\documentclass[twocolumn]{aastex631}
\usepackage{amsmath}
\usepackage{rotating}

\begin{document}

\title{Determining the Host Stars of Planets in Binary Star Systems with Asterodensity Profiling: Investigating the Canonical Radius Gap}

\author[0009-0000-2051-8120]{Nathanael Burns-Watson}
\affiliation{University of Texas at Austin, Department of Astronomy}
\email{nathanael.burnswatson@austin.utexas.edu}

\author[0000-0001-6873-8501]{Kendall Sullivan}
\affil{Department of Astronomy and Astrophysics, University of California Santa Cruz, Santa Cruz CA 95064, USA}
\affiliation{Mullard Space Science Laboratory, University College London, Holmbury St Mary, Dorking, Surrey RH5 6NT, UK}
\email{kendall.sullivan@ucl.ac.uk}

\author[0000-0001-9811-568X]{Adam L. Kraus}
\affiliation{University of Texas at Austin, Department of Astronomy}
\email{alk@astro.as.utexas.edu}



\begin{abstract}
Over the past 30 years, thousands of exoplanets have been discovered, revealing detailed demographics of planets outside the Solar System. One of the most dramatic features of the planet radius distribution is the radius gap, a lack of planets between $\sim$1.8-2 $R_\oplus$. The radius gap is thought to mark the distinction between rocky and gas planets. Recent research has found that the radius gap may not be present in binary star systems. In past studies of planets in binary star systems, the common assumption has been that all of the planets are hosted by the primary star. In many cases, the radius of the planet would be significantly larger if it were orbiting the companion star, which could potentially affect the true radius distribution. It is possible to identify the host stars of planets through stellar density estimates obtained from transit fitting. Using this method, we made probabilistic estimates for the host stars of a sample of 15 transiting exoplanets across 10 binary star systems hosting either 1 or 2 planets, at least one of which would reside in the canonical radius gap if it was circumprimary. We found that 5 of the planets are highly likely to be circumprimary, while the remainder have ambiguous host stars. The lack of unambiguously circumsecondary planets is caused by physical and observational biases that favor circumprimary planets. Nonetheless, the summed posterior probabilities suggest that the canonical radius gap appears less vacant for planets in binaries.
\end{abstract}



\section{Introduction} \label{sec:back}
There have been over 6,000 exoplanets discovered to date, and that number continues to increase with time \citep{ExoplanetArchive}\footnote{Exoplanet Archive database accessed on July 28, 2025: https://exoplanetarchive.ipac.caltech.edu/index.html}. This large population of known planets has made it possible to identify trends in the properties of exoplanets. One of these trends is the radius gap, an observed feature of exoplanet populations whereby
there are comparatively very few planets with radii in the range of $\sim1.5-2.0 R_\oplus$ \citep[e.g.][]{Fulton_2017, VanEylen_2018, Petigura_2022, Ho_2023}. The radius gap is physically meaningful, marking the distinction between smaller Earth-like terrestrial planets and large Neptune-like gaseous planets \citep{Owen_2013}. Several physical processes have been suggested that could cause the radius gap. One of the most prominent is photoevaporation, where the far-UV and X-ray flux of the host star heats the atmosphere to high temperatures, causing atmospheric escape on planets without a sufficiently massive core \citep[e.g.][]{Lammer_2003, Lopez_2013, Owen_2013}. Another common hypothesis is core-powered mass loss, where the bolometric luminosity of the host star combined with a young planet's remnant thermal energy heats the atmosphere and drives escape, but at a longer timescale compared to photoevaporation \citep[e.g.][]{Ginzburg_2018, Gupta_2020}. Other models such as gas-poor formation and collisional evaporation have also been suggested to explain the radius gap \citep[e.g.][]{Lopez_2018, Lee_2021, Chance_2022, Ho_2024}.

Most previous exoplanet research has focused on planets hosted in single-star systems. However, half of all Sun-like stars in the Milky Way are part of binary star systems \citep{Raghavan_2010}, and a stellar companion can have significant effects on planet formation \citep[e.g.][]{Cieza_2009, Kraus_2012, Harris_2012, Barenfeld_2019} and/or survival  \citep[e.g.][]{Kraus_2016, Hirsh_2021, Ziegler_2021, Fontanive_2024}. Many exoplanet hosts have been found to be binary stars \citep[e.g.][]{Kraus_2016, Furlan_2017, Ziegler_2017, Ziegler_2018}. From an observational perspective, unresolved binaries affect our interpretation of transiting planet lightcurves, including the inferred radius and insolation of the planet \citep[e.g.][]{Ciardi_2015, Sullivan_2022}. Additionally, in the search for planets similar to Earth, promising candidates have been found in multiple star systems such as the $\alpha$ Centauri triple system \citep{Anglada_2016}, and many of the best stars to search for other Earths are actually binary star systems.

Research on the radius distribution of planets in binary systems has found that the canonical radius gap may be filled in \citep{Sullivan_2023}.
It may be the case that the radius gap is still present, but its location shifts as a function of binary separation \citep{Sullivan_2024}. \citet{Sullivan_2024} found that close binaries in particular might have a fundamentally different balance of Sub-Neptunes versus Super-Earths. Either case has serious implications for how planets form in both single and binary star systems. One possibility is that the presence of another nearby star decreases the typical core masses of planets that form. This would preserve the radius gap, but it would become a function of the separation between the two stars. In a sample spanning a wide range of binary separations, the radius gap would get averaged out \citep{Sullivan_2023, Sullivan_2024}. Alternatively, the core mass distribution could still be the same, but since circumstellar disk lifetimes are shorter in binaries, there may not be enough time for planets to accumulate substantial atmospheres \citep{Cieza_2009, Kraus_2012, Barenfeld_2019}. This would also lead to a separation-dependent radius distribution.

However, the host stars of these planets are unknown because close binaries are spatially unresolved in data from most transit-search telescopes (including Kepler and TESS). Like other studies of planets in binaries, the \citet{Sullivan_2023} analysis assumed that all the planets were orbiting the primary star in each system. This is because the flux dilution is less severe for the primary star, making it statistically more likely to detect planets around the primary \citep{Gaidos_2016}. This assumption might fail for any given planet. For a planet that is actually orbiting the secondary star, the true radius value would be larger than what was inferred. This is because the transit depth around a fainter star would be proportionally larger which would require a physically larger planet \citep{Ciardi_2015}. It is also unclear whether planets in binary systems mostly orbit the primary star, the secondary star, or a mix of both.

Therefore, it is necessary to identify the host star to measure accurate planet properties. One method for doing so relies on the asterodensity profiling technique, whereby the density of a planet's host star can be estimated from transit fitting \citep{Seager_2003, Kipping_2014}. If there are independent estimates of the stellar densities in a binary, than the asterodensity estimate should be consistent with the star that is hosting the planet \citep[e.g.][]{Sliski_2014, Barclay_2015, Torres_2015, Gaidos_2016, Lester_2022, Eastman_2023}. Other methods for determining the host star include spatially resolving the binary while observing transits \citep[e.g.][]{Leger_2009}, radial velocity follow-up \citep[e.g.][]{Furlan_2017b}, transit timing variations \citep[e.g.][]{Montalto_2010}, and astrometric offsets during transits \citep[e.g.][]{Hadjigeorghiou_2024}.

Planets form from protoplanetary disks of material that surround young stars, eventually coalescing into distinct objects. How is planet formation in these disks affected when another large gravitational potential is nearby? Are the building blocks of these planets perturbed? Does it affect the formation of planetary atmospheres? The radius distribution and host star frequencies for planets in binary systems would provide valuable tests of theories regarding the formation and physical composition of exoplanets.

\section{Observations}
The stellar parameters of the Kepler target stars were originally estimated from 2MASS, and other optical imaging \citep{Latham_2005, Batalha_2010, Brown_2011}. They were later updated with wide variety of photometric and spectroscopic data including Kepler itself, the Large Sky Area Multi-Object Fiber Spectroscopic Telescope, and the Apache Point Observatory for Galactic Evolution Experiment \citep{Mathur_2017}. The stellar parameters have been further updated and made more precise using Gaia \citep{Berger_2018, Berger_2023}, but the stellar parameters from \citet{Mathur_2017} were still widely used while the key results of Kepler were being finalized.

The stellar parameters in \cite{Mathur_2017} assumed that all Kepler targets were single stars, but some of the Kepler targets were unresolved binaries. To make accurate estimates of the planetary properties in unresolved binaries, \citet{Sullivan_2022} used the results from previous high-resolution imaging combined with unresolved spectra of the binaries from the Hobby-Eberly Telescope to estimate the physical parameters for both stars in a given binary. They specifically selected for unresolved Kepler binaries that hosted planets and planetary candidates with initial radius estimates $<$2.5$R_\oplus$. They estimated flux contamination correction factors for each star \citep{Ciardi_2015, Furlan_2017}:
\begin{equation} \label{eq:f_cor}
    \begin{split}
        f_{pri} & = \sqrt{1 + 10^{-0.4\Delta m}} \\
        f_{sec} & = \sqrt{1 + 10^{+0.4\Delta m}}
    \end{split} 
\end{equation}

Where $\Delta m$ is the contrast between the stars in magnitudes of the band where the lightcurves are taken. These correction factors were be used to estimate the radius of the planet if it is orbiting either the primary or secondary star \citep{Sullivan_2023}:
\begin{equation} \label{eq:R_cor}
    \begin{split}
        R_{p,pri} & = R_{p,old} * \frac{R_{\star,pri}}{R_{\star,old}} * f_{pri} \\
        R_{p,sec} & = R_{p,old} * \frac{R_{\star,sec}}{R_{\star,old}} * f_{sec}
    \end{split} 
\end{equation}

Where $R_p$ is the planet's radius, $R_{\star}$ is the star's radius, $old$ refers to the original estimates that assumed single stars, and $pri$ and $sec$ refer to the primary and secondary star estimates from the \citet{Sullivan_2023} analysis, respectively.

We used the \citet{Sullivan_2023} catalog of planets in binary star systems and selected all targets with a period $\log(P) < 1.5$, or $P < 31.6$ days. This ensured that the targets would have numerous well-sampled transits over Kepler's 4-year mission. We selected targets that, if they are orbiting the primary star, would be within $\pm 0.2 R_{\oplus}$ using the functional form of the radius gap from the \citet{Petigura_2022} 0.7-1.0 $M_\oplus$ sample:
\begin{equation}
    R_{p,gap} = 1.7 R_\oplus \left(\frac{P}{10days} \right)^{-0.13}
\end{equation}
We also selected binary systems that contain no more than 2 total planets or planetary candidates. Systems with more than 2 planets were too computationally expensive for this analysis, but future work will look at binaries with higher-order planet multiplicity.
 
\begin{figure}
    \centering
    \includegraphics[width=1\linewidth]{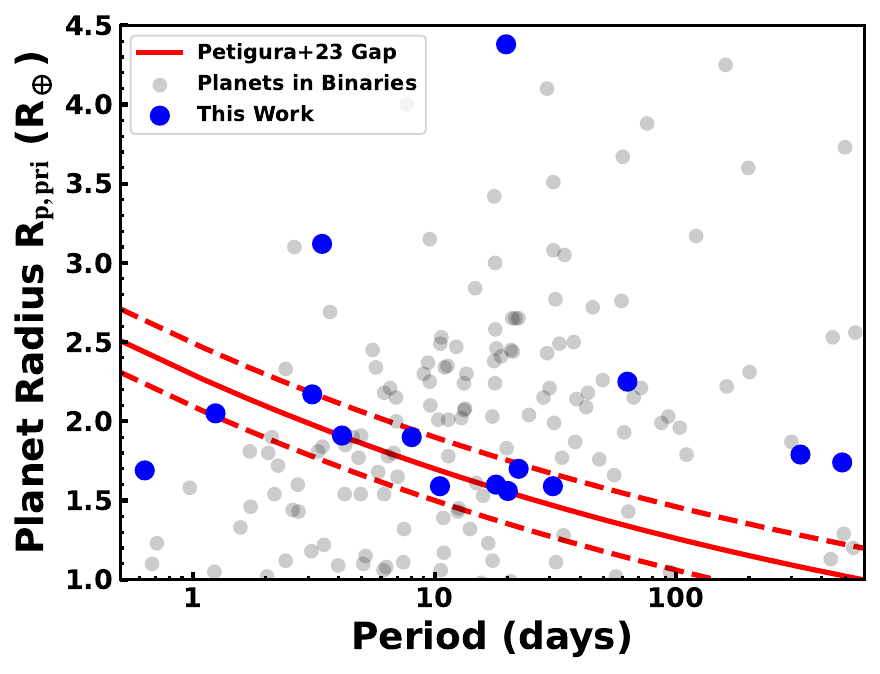}
    \caption{$R_{p,pri}$ from \citet{Sullivan_2023} versus orbital period for planets in binaries. The gray points show the planets hosted in binary star systems from the \citet{Sullivan_2023} catalog. The solid red line shows the functional form of radius gap as defined by \citet{Petigura_2022}. The dashed red lines are $\pm 0.2 R_\oplus$ of the functional radius gap. The gray points within the radius gap range have not been analyzed in this work because they have more than 2 planets/planetary candidates in the system or because the quality of the system fit was poor. The blue points are the planets that have been analyzed in this work. This sample focuses on planets that would reside in the canonical radius gap if they were circumprimary.}
    \label{fig:rvp}
\end{figure}

The aim of these criteria was to see how many of the planets in our sample might actually be orbiting the secondary star, and if this was sufficient for a gap to reappear. There are a total of 18 systems in the \citet{Sullivan_2023} catalog that satisfy these criteria. 9 of these have been eliminated from our sample. KOIs 1973, 2486, and 3112 were removed because they are possible triple star systems. KOIs 2179 and 6475 were removed because at least one of the stars is too intrinsically variable to obtain good fitting results without additional lightcurve detrending. KOIs 270, 4768, and 4823 were flagged because the  \cite{Sullivan_2023} analysis did not result in a good fit to the data. This means that the independent density estimates for the stars in these systems are unreliable. KOI-1101 was flagged for future investigation because it may contain either 1 or 2 planets in its system and it requires further analysis. In this work, we present the results for KOIs 1700, 1899, 2289, 2580, 2851, 3120, 3401, 3456, and 5845. The details of this final sample are shown in Table \ref{tab:sample} and Figure \ref{fig:rvp}. There are a total of 10 KOIs (including KOI-1300, discussed later) with 15 total planets or planetary candidates, 9 of which are in the radius gap subsample.

We analyzed an additional system, KOI-1300, as a test of our methodology. KOI-1300.01, the only known planet in the system, would have a radius of $1.69 R_\oplus$ if it is orbiting the primary star. It is not within our defined radius gap range because of its short orbital period ($\sim0.6$ days), but it is close. However, this short period also means that the transits are well sampled, making it a good case to test our methodology. Additionally, this planet has already been confirmed, but it was thought to only be a single star system at the time \citep{Johnson_2017}. This means that a correction to the planet's canonical parameters is needed.

\begin{deluxetable*}{lccccccccc}
    \tabletypesize{\scriptsize}
    \tablewidth{0pt} 
    \tablecaption{Previous estimates for the planetary and stellar parameters of the systems analyzed in this work \label{tab:sample}}
    \tablehead{
    \colhead{KOI} & \colhead{Period} & \colhead{$\mathrm{R_{p,pri}}$} & \colhead{$\mathrm{R_{p,sec}}$} & \colhead{$\mathrm{m_{Gaia}}$} & \colhead{Projected Separation} & \colhead{$\mathrm{T_{eff,pri}}$} & \colhead{$\mathrm{T_{eff,sec}}$} & \colhead{$\mathrm{\rho_{pri}}$} & \colhead{$\mathrm{\rho_{sec}}$} \\
    \colhead{} & \colhead{[days]} & \colhead{[$\mathrm{R_\oplus}$]}& \colhead{[$\mathrm{R_\oplus}$]} & \colhead{[mag]} & \colhead{[arcsec]} & \colhead{[K]} & \colhead{[K]} & \colhead{[$\mathrm{g/cm^3}$]} & \colhead{[$\mathrm{g/cm^3}$]}
    } 
    \colnumbers
    \startdata 
    1300.01 & 0.63   & ${1.69}^{+0.17}_{-0.18}$ & ${3.40}^{+0.37}_{-0.37}$  & 14.285 & 0.75  & ${4955}^{+56}_{-56}$  & ${3930}^{+31}_{-38}$   & ${3.39}^{+0.05}_{-0.05}$ & ${6.00}^{+0.11}_{-0.14}$ \\
    \hline
    1700.01* & 8.04   & ${1.90}^{+0.32}_{-0.34}$  & ${2.71}^{+0.52}_{-0.52}$ & 14.471 & 0.282 & ${5467}^{+97}_{-103}$ & ${4756}^{+121}_{-125}$ & ${2.23}^{+0.06}_{-0.07}$ & ${3.10}^{+0.11}_{-0.11}$ \\
    \hline
    1899.01 & 19.76  & ${4.38}^{+0.94}_{-2.48}$ & ${4.29}^{+2.19}_{-1.89}$ & 14.563 & 1.844 & ${6094}^{+45}_{-48}$  & ${5498}^{+52}_{-64}$   & ${1.14}^{+0.02}_{-0.01}$ & ${1.43}^{+0.02}_{-0.02}$ \\
    1899.02* & 10.52  & ${1.59}^{+0.47}_{-0.85}$ & ${1.75}^{+0.74}_{-0.83}$ & 14.563 & 1.844 & ${6094}^{+45}_{-48}$  & ${5498}^{+52}_{-64}$   & ${1.14}^{+0.02}_{-0.01}$ & ${1.43}^{+0.02}_{-0.02}$ \\
    \hline
    2289.01 & 62.78  & ${2.25}^{+0.51}_{-0.52}$ & ${8.70}^{+2.05}_{-2.10}$   & 13.358 & 0.94  & ${6370}^{+44}_{-40}$  & ${3838}^{+45}_{-43}$   & ${1.00}^{+0.01}_{-0.01}$  & ${6.20}^{+0.22}_{-0.21}$  \\
    2289.02* & 20.1   & ${1.56}^{+0.37}_{-0.38}$ & ${6.03}^{+1.43}_{-1.45}$ & 13.358 & 0.94  & ${6370}^{+44}_{-40}$  & ${3838}^{+45}_{-43}$   & ${1.00}^{+0.01}_{-0.01}$  & ${6.20}^{+0.22}_{-0.21}$  \\
    \hline
    2580.01* & 3.12   & ${2.17}^{+0.71}_{-0.86}$ & ${2.84}^{+0.98}_{-1.06}$ & 15.790  & 0.60   & ${5643}^{+71}_{-74}$  & ${4961}^{+65}_{-63}$   & ${1.97}^{+0.04}_{-0.04}$ & ${2.87}^{+0.05}_{-0.05}$ \\
    \hline
    2851.01 & 3.42   & ${3.12}^{+0.47}_{-0.49}$ & ${3.26}^{+0.51}_{-0.51}$ & 15.472 & 0.39  & ${5837}^{+139}_{-98}$ & ${5720}^{+154}_{-117}$ & ${1.68}^{+0.07}_{-0.05}$ & ${1.75}^{+0.08}_{-0.06}$ \\
    2851.02* & 1.24   & ${2.05}^{+0.34}_{-0.34}$ & ${2.15}^{+0.35}_{-0.35}$ & 15.472 & 0.39  & ${5837}^{+139}_{-98}$ & ${5720}^{+154}_{-117}$ & ${1.68}^{+0.07}_{-0.05}$ & ${1.75}^{+0.08}_{-0.06}$ \\
    \hline
    3120.01* & 4.14   & ${1.91}^{+0.45}_{-0.43}$ & ${2.20}^{+0.48}_{-0.51}$  & 14.826 & 1.14  & ${6504}^{+60}_{-55}$  & ${6066}^{+58}_{-53}$   & ${0.90}^{+0.02}_{-0.02}$  & ${1.58}^{+0.04}_{-0.03}$  \\
    \hline
    3401.01* & 17.96  & ${1.60}^{+0.32}_{-0.32}$  & ${4.57}^{+0.92}_{-0.93}$ & 14.388 & 0.65  & ${5809}^{+72}_{-64}$  & ${3963}^{+53}_{-35}$   & ${1.75}^{+0.04}_{-0.03}$ & ${5.43}^{+0.17}_{-0.12}$ \\
    3401.02 & 326.67 & ${1.79}^{+0.65}_{-0.65}$ & ${5.08}^{+1.83}_{-1.86}$ & 14.388 & 0.65  & ${5809}^{+72}_{-64}$  & ${3963}^{+53}_{-35}$   & ${1.75}^{+0.04}_{-0.03}$ & ${5.43}^{+0.17}_{-0.12}$ \\
    \hline
    3456.01* & 30.86  & ${1.59}^{+0.27}_{-0.28}$ & ${1.61}^{+0.28}_{-0.29}$ & 12.988 & 0.044 & ${5680}^{+53}_{-56}$  & ${5655}^{+62}_{-74}$   & ${2.12}^{+0.03}_{-0.03}$ & ${2.28}^{+0.04}_{-0.04}$ \\
    3456.02 & 486.13 & ${1.74}^{+0.67}_{-0.69}$ & ${1.75}^{+0.69}_{-0.69}$ & 12.988 & 0.044 & ${5680}^{+53}_{-56}$  & ${5655}^{+62}_{-74}$   & ${2.12}^{+0.03}_{-0.03}$ & ${2.28}^{+0.04}_{-0.04}$ \\
    \hline
    5845.01* & 22.27  & ${1.7}^{+0.47}_{-0.5}$   & ${1.68}^{+0.49}_{-0.51}$ & 11.299 & 1.65  & ${6850}^{+34}_{-39}$  & ${6878}^{+35}_{-45}$   & ${0.65}^{+0.01}_{-0.01}$ & ${0.68}^{+0.01}_{-0.01}$
    \enddata
\tablecomments{Planetary and stellar parameters are taken from \citet{Sullivan_2023}. Planets with * are part of the radius gap subsample.}
\end{deluxetable*}

\section{Analysis/Methodology}
\subsection{Lightcurve Assessment}
We accessed the Kepler lightcurves for each target using \emph{lightkurve} \citep{lk}, which we also used to normalize the lightcurve and remove systematics using the default parameters. We used the pre-search data conditioning simple aperture photometry (PDCSAP) flux values for our analysis. The PDCSAP lightcurves have the systematic artifacts removed, making them better suited for studying planets. We selected for 30-minute cadence data and used the full time span of data available for each system. A comparison of the results obtained with 30-minute versus 60-second data can be found in Section~\ref{30v60}.

\subsection{Lightcurve Fitting}
To generate and fit transit models, we used \emph{exoplanet} \citep{exoplanet} and \emph{pymc3} \citep{pymc3}. The parameters in our model are the mean out of transit flux of the lightcurve ($\mu$), the period of the planet ($P$), a reference transit time ($t_0$), 2 quadratic limb darkening parameters ($u_0$, $u_1$), the radius of the planet relative to the radius of the host star ($r/R_\star$), the density of the host star ($\rho_\star$), and the impact parameter of the transit ($b$). We used a Gaussian prior distribution for $\mu$ with a mean of 1 and a standard deviation of 0.001. We used a  Gaussian for $\log(P)$ with a standard deviation of 0.1 and then converted back to linear $P$. We used the reported orbital period on Exoplanet Archive as the initial guess for the period \citep{ExoplanetArchive}. We used uniform prior distributions for $t0$, $r/R_\star$, and $\rho_\star$. We used specialized prior distributions from \emph{exoplanet} for $u$ and $b$ \citep{exoplanet}. The priors for the limb darkening parameters are derived in \citet{Kipping_2013}. The prior for the impact parameter is derived in \citet{Kipping_2016} and is of the form $p \propto (1-b^2)^{1/4}$ where $p$ is the probability distribution. Since the target planets in our sample only consisted of short period planets in binaries (see Table \ref{tab:sample}), we expected eccentricities to be low, so all orbits were assumed to be circular \textbf{\citep[e.g.][]{Kipping_2013b, VanEylen_2019}}. We tested models that incorporated eccentricity and argument of periastron and found that they did not substantially change the posteriors, but they did increase run times. Unless the eccentricity of a planet's orbit is very high, fitting a circular orbit will not introduce significant error to the inferred stellar host density \citep{Kipping_2014}.

\begin{deluxetable*}{lccc}
    \tabletypesize{\scriptsize}
    \tablewidth{0pt} 
    \tablecaption{Priors for Our Transit Model \label{tab:priors}}
    \tablehead{
    \colhead{Parameter} & \colhead{Range} & \colhead{Description}} 
    \startdata
    Mean & Gaussian with $\mu$=1.000, $\sigma$=0.001 & Mean out of transit normalized flux \\
    Period & Lognormal & Planet's orbital period \\
    Reference Time & Uniform with range of Period & Time of a reference transit \\
    Limb Darkening & \citet{Kipping_2013} parameterization & Quadratic limb darkening model \\
    Radius Ratio & Uniform ($\pm1$ dex of initial guess) & Ratio of planet radius to host star radius \\
    Stellar Density & Uniform (0-25 g/cm$^3$) & Host star density \\
    Impact Parameter & \citet{Kipping_2016} parameterization & Impact parameter of the transit
    \enddata
    \tablecomments{The parameters and priors used in our transit model. The first column names the model parameter. The second column defines the functional form (and range, where relevant) of the prior for the parameter. The third column provides a brief description of the parameter.}
\end{deluxetable*}

The PDCSAP fluxes are already corrected for contamination from known non-target sources. However, the binaries in this work are within 2" separation and were not known when these initial corrections were applied. This is reflected in the Crowding SAP values - the fraction of flux thought to be coming from the target star - which are all $>$0.99 for the systems analyzed in this work. Using flux corrections from \cite{Sullivan_2023}, we corrected the lightcurves for the contamination around the primary and secondary stars. This correction was of the form:
\begin{equation} \label{eq:flux_cor}
    \begin{split}
        \mathcal{L}_{pri} & = f_{pri} (\mathcal{L}_0 - 1) + 1 \\
        \mathcal{L}_{sec} & = f_{sec} (\mathcal{L}_0 - 1) + 1 
    \end{split} 
\end{equation}

Where $\mathcal{L}_0$ is the uncorrected lightcurve (scaled to an out of transit flux of 1), $f_{pri}$ and $f_{sec}$ are defined in Equation \ref{eq:f_cor}, and $\mathcal{L}_{pri}$ and $\mathcal{L}_{sec}$ are the contamination corrected lightcurves for the primary and secondary stars, respectively. This adjusted the transit depth to what it would be if the planet was orbiting the primary or secondary star by accounting for the flux contamination and making the transit deeper. To perform the transit fit and measure revised planetary orbital properties, we fit a transit model to the contamination-corrected lightcurves.

We used the reported values on Exoplanet Archive as initial guesses for a transit model represented by the parameters described previously. We then ran an initial optimization routine from \emph{pymc3} to find a best-fit model. With this best-fit model established, we used a No U-Turn Sampler (NUTS) \citep{HMC}, a type of Markov Chain Monte Carlo (MCMC) method, to sample from the parameter space and obtain posterior distributions for all of our model parameters. The benefit of the NUTS method is that it avoids random walks and uses gradient information to sample the parameter space more efficiently than other MCMC methods like the Metropolis-Hastings algorithm \citep{Metro_1953}. This is particularly useful for models that contain large numbers of parameters like the transit model for our data. 

After performing the transit fit and measuring posteriors for all relevant parameters, we used the posterior distributions of the model parameters to estimate the density posterior of the host star using the equation from \citet{Kipping_2014}:
\begin{equation}
    \rho_\star = \frac{3 \pi (a/R_\star)^3}{G P^2}
\end{equation}

\subsection{Host Star Identification}
We identified the host stars by comparing the measured stellar density from the transit fit to the independent, spectroscopic estimates for the host star density from \citet{Sullivan_2023}. By comparing the densities, we can calculate a probability for the host star. The densities that agree between the two methods indicate which star is the likely host. Similarly, the densities should not agree for the star that is not hosting the planet. This method, known as asterodensity profiling, was first demonstrated in \citet{Kipping_2014} and has been utilized in several papers since \citep[e.g.][]{Sliski_2014, Torres_2015, Gaidos_2016, Lester_2022, Eastman_2023}. To determine the probability of a planet being hosted by either star, we applied Bayes' Theorem:
\begin{equation}
    P(A|B) = P(A) * \frac{P(B|A)}{P(B)}
\end{equation}

In this case, $A$ has 2 possibilities - the planet actually being circumprimary or the planet actually being circumsecondary, and $B$ is the observation. $P(A|B)$ is referred to as the posterior probability, $P(A)$ is the prior probability, $P(B|A)$ is the likelihood, and $P(B)$ is the marginal probability.

For our prior probability $P(A)$, we modeled the single-star planet radius distribution from \citet{Fulton_2017} as a generalized gamma function. We used the observed planet radii and associated completeness weights from \citet{Fulton_2017} to create a simulated population of $10^6$ planets that represents that true underlying distribution of planet radii. We used maximum likelihood estimation (MLE) to fit a generalized gamma distribution to this simulated population of planets. As part of the MLE, we masked any data points in the radius gap range of 1.5-2.0 $R_\oplus$. This ensured that our prior was agnostic to the presence and structure of the radius gap. The full functional form of our prior with best fit parameters is: 
\textbf{\begin{align}\label{eq:raddist}
    &P(R_p) = A \frac{|c|(x-B)^{ca-1}\exp((x-B)^c)}{\Gamma(a)} \\
    A = &0.267 \quad B = 3.18 \quad a = 1.17 \quad c = 2.16 \nonumber
\end{align}
}Where $\Gamma$ is the gamma function.

\begin{figure}
    \centering
    \includegraphics[width=1\linewidth]{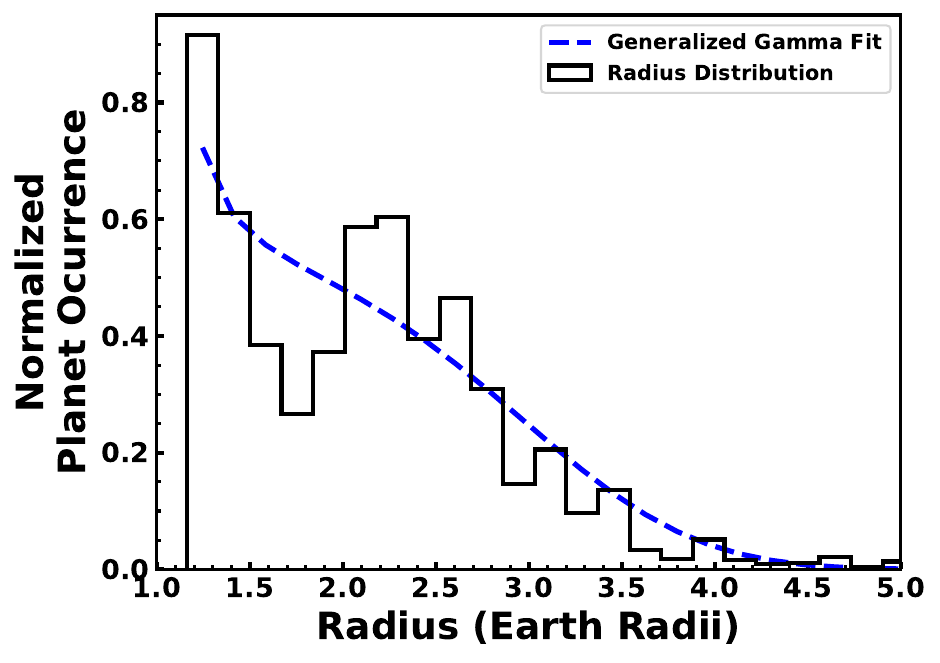}
    \caption{The radius distribution prior that was used in this work. The black histogram is the completeness-corrected single planet radius distribution from \citet{Fulton_2017}. The dashed blue is the model radius distribution that was used as our prior. The model fit did not include planets in the range of 1.5-2.0 R$_\oplus$, making the prior agnostic to the presence of the radius gap. A prior that included the radius gap would differ by a factor of $\sim$2 within the radius gap range. The model used in this work imposes the constraint that small planets are more numerous than large planets and are therefore more probable to occur. Since a planet's inferred radius is different depending on the host star, we used this to assign prior probabilities of a planet being circumprimary versus circumsecondary.
    \label{fig:prior}}
\end{figure}

Since planets with smaller radii are more numerous, our prior favors smaller radius planets as being more probable. The planets in \citet{Sullivan_2023} almost always have $R_{p,sec}$ that is larger than $R_{p,pri}$. This means that our prior systematically favors primary star hosts. The conservative approach is to not include the radius gap in our prior, and then see if it re-emerges. A prior that favors circumprimary planets also naturally folds in the observational bias that we are more likely to detect planets around the primary star due to flux contamination decreasing the fractional transit depth.

For our likelihood $P(B|A)$, we measured the overlap of the densities estimated from the transit fitting with the spectroscopic density of each star, as we show in Figure \ref{fig:1300} (bottom). We did this by multiplying the transit posterior distribution with the spectroscopic posterior distribution for each outcome $B$. We used the relative overlaps between the primary and secondary stars to normalize the likelihoods such that they summed to unity.

The marginal probability $P(B)$ cannot be easily calculated beforehand, so it was determined afterward such that the posterior probabilities $P(A|B)$ of the two possibilities (circumprimary or circumsecondary) summed to unity.

\section{Results}
\subsection{KOI-1300}

\begin{figure*}[ht]
    \gridline{\includegraphics[angle=-90, width=0.6\textwidth]{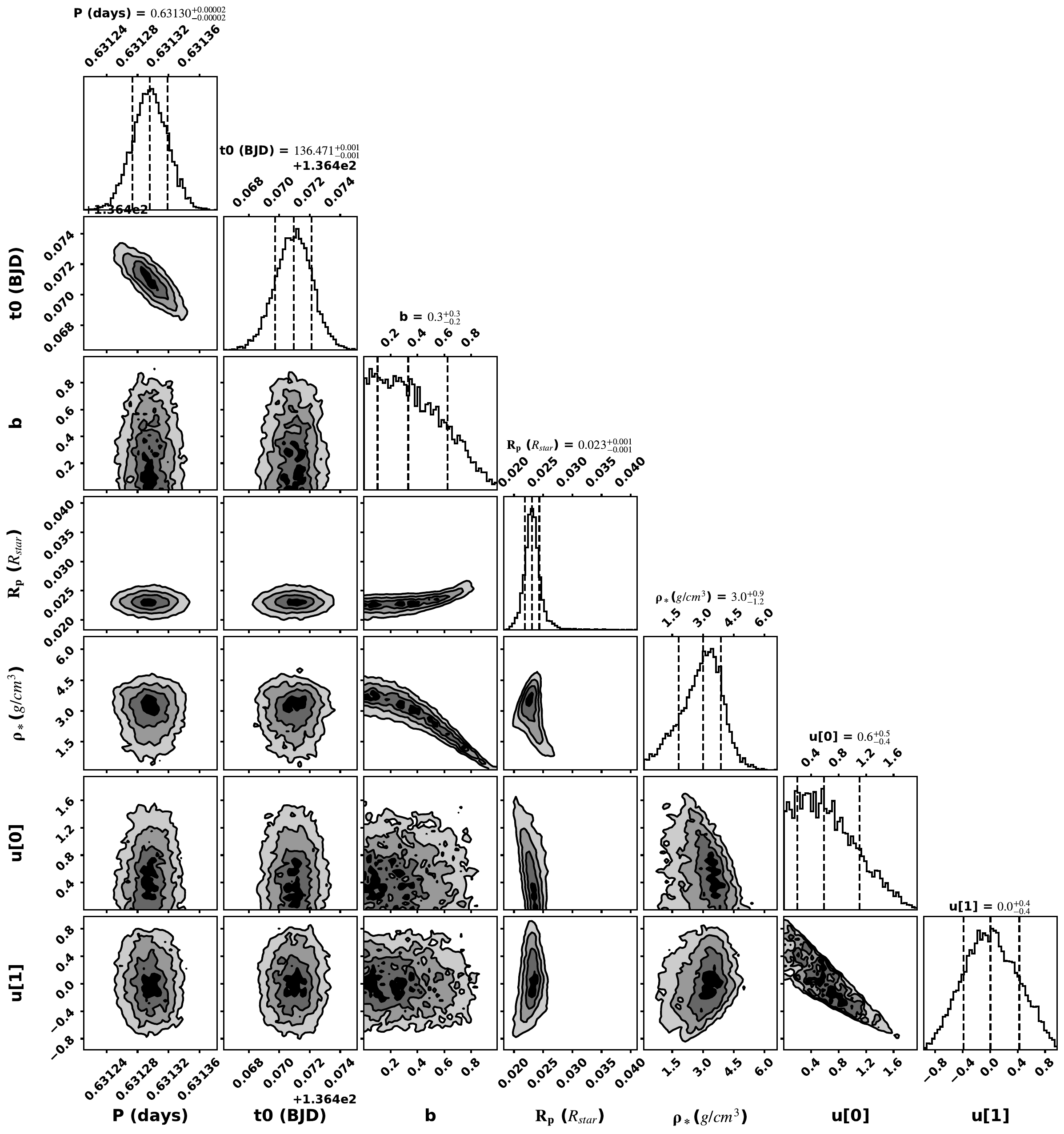}
        \includegraphics[angle=-90, width=0.35\textwidth]{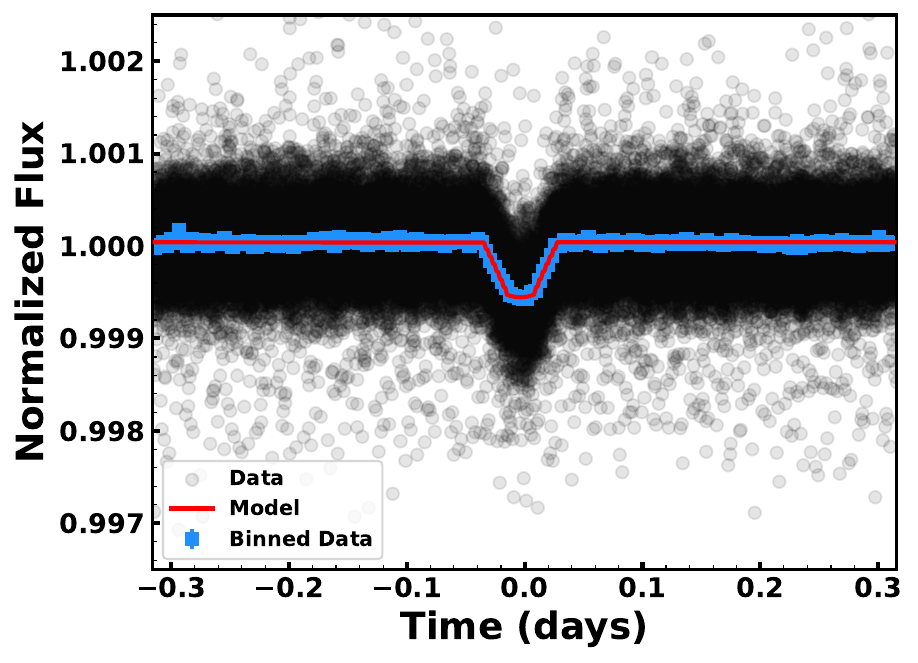}}
    \gridline{\includegraphics[angle=-90, width=0.6\textwidth]{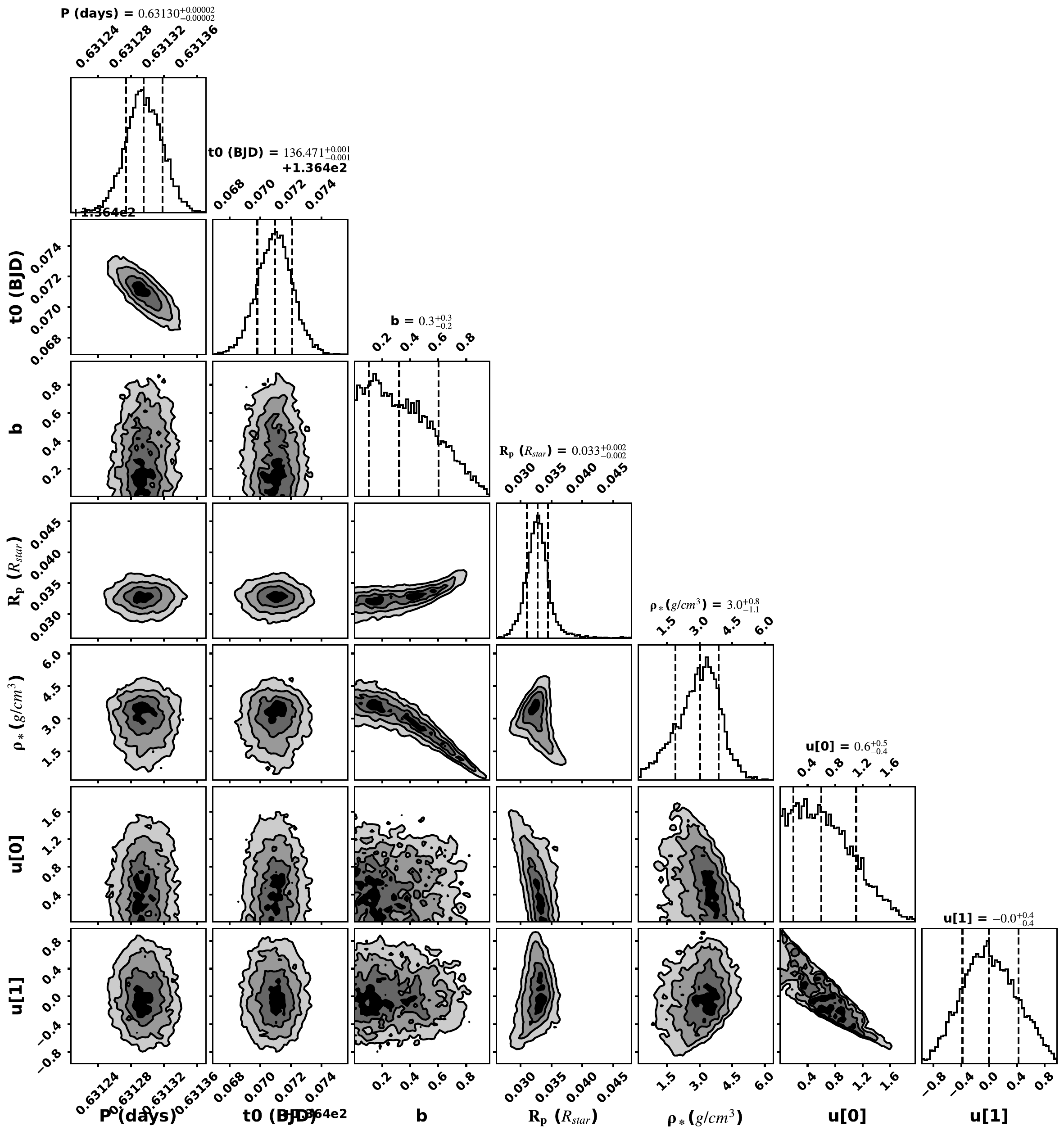}
        \includegraphics[angle=-90, width=0.35\textwidth]{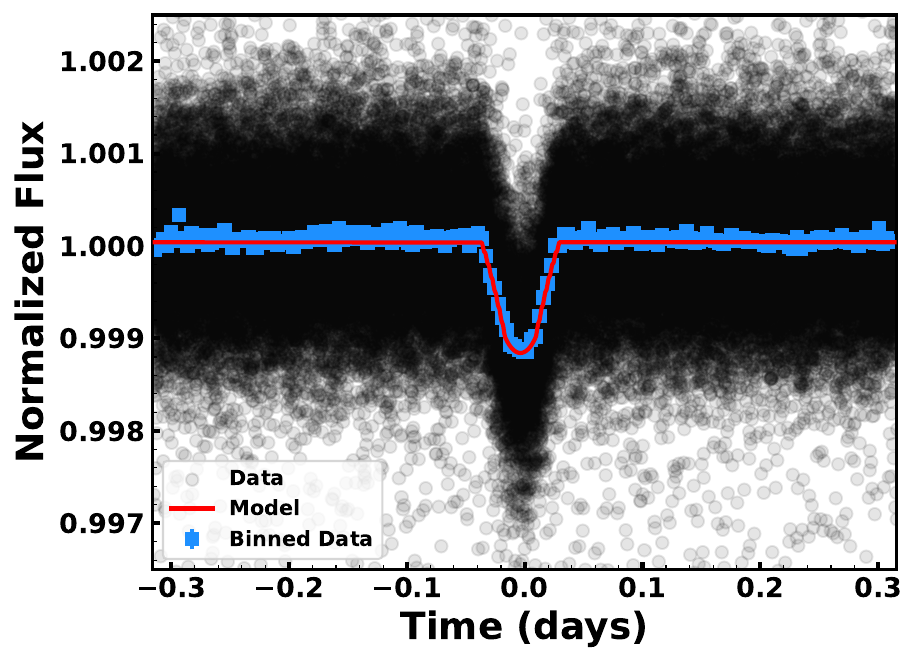}}
    \caption{The transit fitting results for KOI-1300.01. The upper right panel shows the phase-folded and contamination-corrected transit lightcurve for the planet's circumprimary case. The gray circles are the normalized and contamination corrected flux values. The blue squares are the normalized flux values obtained from binning 800 data points at a time. The red line is the model fit to the data. The lower right panel shows the same information for the circumsecondary case. The upper left panels shows the corner plots from the MCMC sampling in the circumprimary case. The lower left shows the corner plots for the circumsecondary case.}
    \label{fig:1300_fit}
\end{figure*}

\begin{figure}
    \centering
    \includegraphics[width=0.45\textwidth]{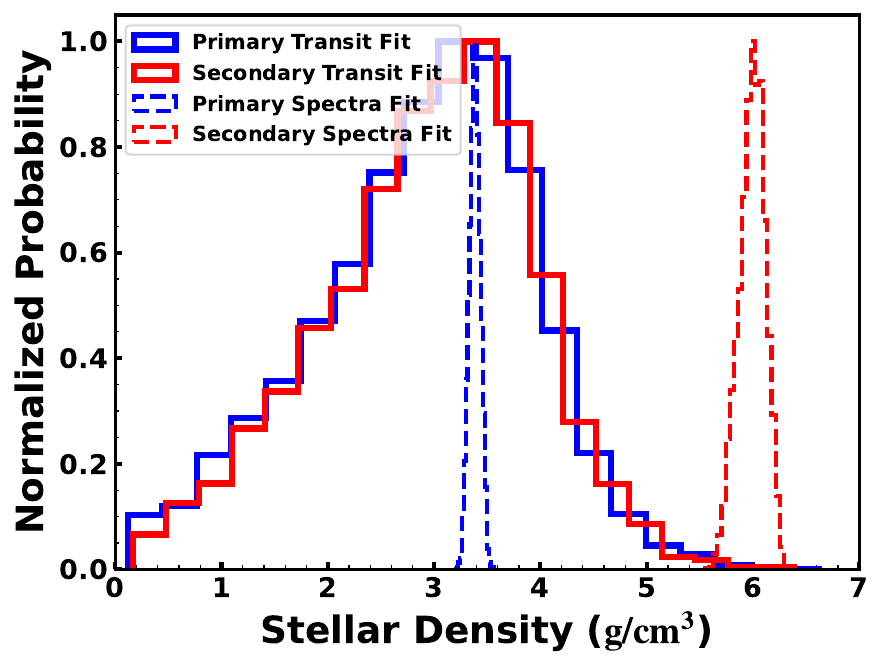}
    \caption{The density posterior distributions for KOI-1300.01. The solid blue histogram shows the posterior density distribution from transit fitting for the primary star case. The solid red histogram shows the same for the secondary star case. The dashed blue and red histograms show the \citet{Sullivan_2023} spectroscopic density distributions for the primary star and secondary star, respectively. Since the asterodensity posterior shows substantial overlap with the primary star but not the secondary star, we consider this planet to be unambiguously circumprimary.}
    \label{fig:1300}
\end{figure}

As a proof of concept for our methodology, we first analyzed the KOI-1300 system, chosen for its deep transit depth and short period. Figure \ref{fig:1300} shows the results we obtained for this system. Panels a and b show the phase-folded lightcurves if the planet orbited the primary and secondary star, respectively. The transit models fit extremely well which can be seen from the the binned data points. Panels c and d are the corner plots from the MCMC sampling.  Panel e shows the final density distributions that we obtained. 

As Figure \ref{fig:1300} shows, the primary density distributions have good agreement, and the secondary density distributions show no overlap at all. This suggests that the planet is highly likely to be orbiting the primary star. The likelihood returns a primary primary probability of 0.995. The prior returns 0.850. This leads to a posterior probability of 0.999. This is a result that unambiguously favors a primary star host.

\subsection{Testing 30-minute Versus 60-second Cadence} \label{30v60}
\begin{figure}
    \centering
    \gridline{\fig{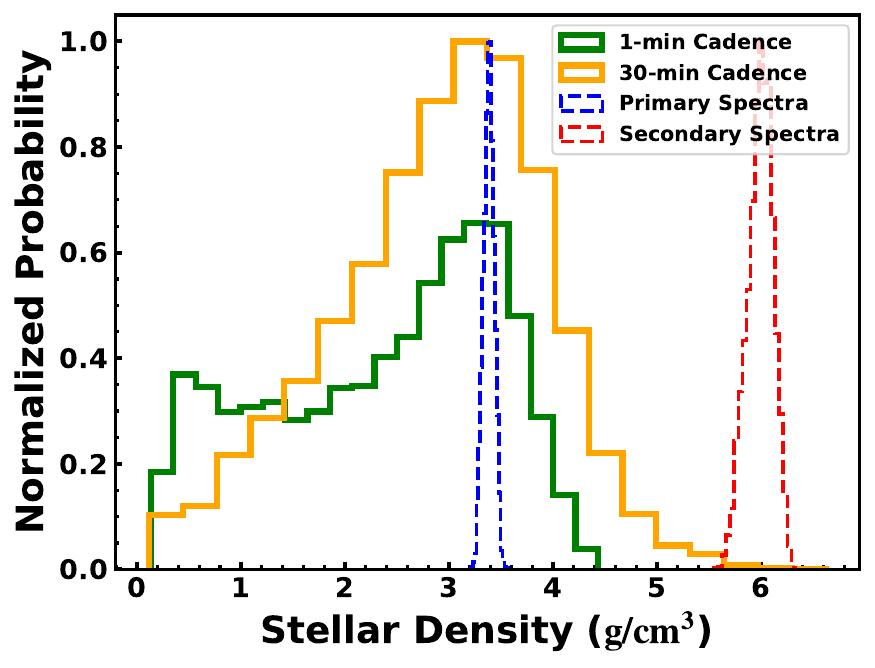}{0.45\textwidth}{}}
    \caption{Comparison of the posterior density distributions for KOI-1300.01 when using 30-minute cadence data versus 60-second cadence data. The solid orange histogram shows the transit density distribution using 30-minute cadence data. The solid green histogram shows the transit density distribution using 60-second cadence data. The dashed blue and red histograms show the \citet{Sullivan_2023} spectroscopic density distributions for the primary star and secondary star, respectively. The results of both are consistent with the planet being circumprimary, but the fast-cadence data yields a broader posterior distribution with a second peak at very low density. This is further discussed in Section \ref{30v60}.}
    \label{fig:60v30}
\end{figure}

An important methodological test was whether there was any difference in the fitting results when using Kepler's 30-minute (slow) cadence data versus the 60-second (fast) cadence data. It seemed plausible that the fast cadence would yield narrower posterior distributions for the model parameters, especially for $b$ and $a/R_\star$ because better sampling frequency would better constrain the shape of the transit. There are significantly fewer systems with fast cadence data, but we were able to test the slow vs fast cadence on our test-case system KOI-1300, which has both fast and slow cadence data. This enabled us to compare the transit fitting in both cases.

For KOI-1300.01, the flux contamination corrections for the lightcurve do not have a strong effect on the posterior density distributions, so they were not applied for this test case. Our results from fitting the transit of KOI-1300.01 using slow versus fast cadence data are shown in Figure \ref{fig:60v30}. The fast-cadence data does not result in a more precise posterior distribution for the stellar density. We concluded that the lower signal-to-noise of the fast-cadence data compared to slow cadence offsets the benefits of higher sampling frequency, resulting in similar posterior distributions. In summary, we decided to use the slow-cadence data for all systems in our analysis because the use of fast-cadence data does not necessarily result in more precise posterior distributions; usinfast-cadence data greatly increases the computational cost of our MCMC sampling; 
and most systems only have slow-cadence data. In the future, it could be valuable to repeat this test on a brighter star with a higher signal-to-noise ratio.

\subsection{Radius Gap Sample}
The overall goal of this work was to perform the host star analysis on planets that would be in the canonical radius gap if they were circumprimary to assess whether enough planets are circumsecondary such that a gap reappears.

We performed the host star analysis on a total of 15 planets, 9 of which are part of the radius gap subsample. The density posteriors for each fitted system are shown in Figures \ref{fig:1planet} and \ref{fig:2planet}. There are many ambiguous results because of the broad posterior distributions, especially in cases where the stellar densities are close in value. In the case of KOI 3456.02, the density posterior is extremely broad because the planet has a long period of about 486 days, meaning that there are not enough transits in the lightcurve to constrain the host star density.

The Bayesian probabilities are shown in Table \ref{tab:results}. The prior probabilities strongly favor ($>0.9$) circumprimary planets for 4 of the total cases. In these cases, the planets show a high circumprimary posterior probability, regardless of the likelihood. This can be seen most clearly in the case of 2289.01, where the likelihood moderately favors the planet to be circumsecondary, but it is outweighed by the extremely high circumprimary prior probability. When the prior probability is near 0.5, it does not have a significant effect on the posterior probability.

\begin{figure}
    \centering
    \gridline{\fig{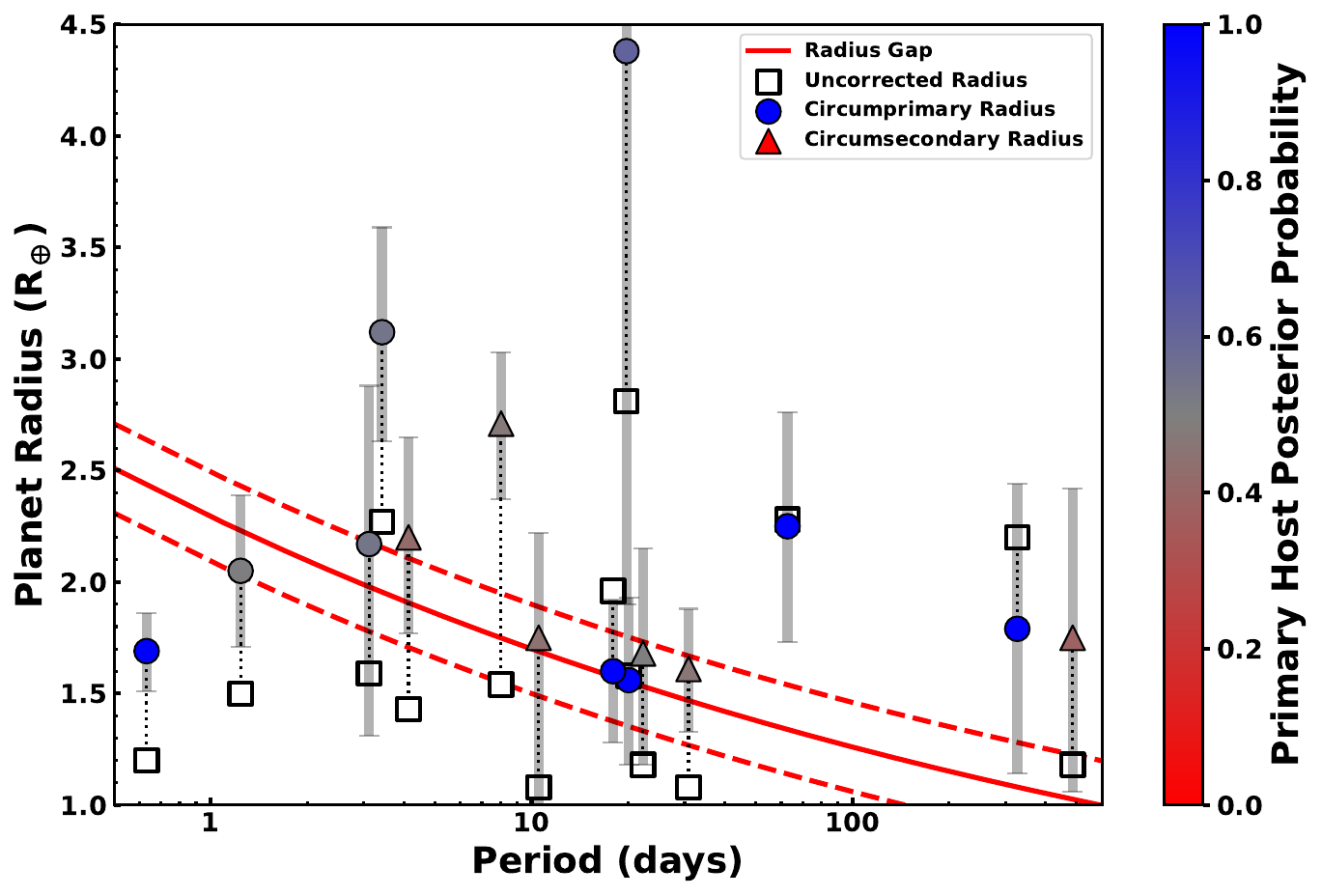}{0.45\textwidth}{}}
    \caption{Radius versus period for the planets analyzed in this work. The empty squares are the uncorrected radii reported in Kepler DR25 \citep{Thompson_2018}. The closed points are the corrected radii, color-coded by primary host posterior probability. Planets with a higher circumprimary probability are marked with circles, and the circumprimary radius is used as the corrected radius. Planets with a higher circumsecondary probability are marked with triangles, and the circumsecondary radius is used as the corrected radius. The radius gap is still occupied. These results do not support the presence of a radius gap for planets in binary star systems.}
    \label{fig:cor_rad}
\end{figure}

The resulting Radius v Period plot is shown in Figure \ref{fig:cor_rad}. There are 5 total planets (KOIs 1300.01, 2289.01, 2289.02, 3401.01, 3401.02) that have a posterior probability $\ge0.95$ of being circumprimary. Two of these (2289.02, 3401.01) are part of the radius gap subsample. We consider these planets to be unambiguously circumprimary. The other 10 planets have posterior probabilities $>0.3$ and $<0.7$ to be circumprimary. We consider these planets to have ambiguous host stars. If we take any planet with a posterior probability $>0.5$ to be circumprimary and $<0.5$ to be circumsecondary, then 9 of the planets would be circumprimary and 6 would be circumsecondary. For planets in the radius gap subsample, 5 would be circumprimary and 4 would be circumsecondary. If we sum the probabilities for all planets, we find that there should be 8.08 circumprimary planets out of 15 from the likelihoods, 10.10 from the priors, and 9.89 from the posteriors.

\begin{deluxetable*}{lccccc}
\tabletypesize{\scriptsize}
\tablewidth{0pt} 
\tablecaption{Statistical Results \label{tab:results}}
\tablehead{
\colhead{KOI} & \colhead{Primary} & \colhead{Prior} & \colhead{Posterior} & \colhead{Updated} \\[-8pt] \colhead{} & \colhead{Likelihood} & \colhead{Probability} & \colhead{Probability} & \colhead{Radius (R$_\oplus$)} & \colhead{Flag}}
\colnumbers
\startdata
1300.01 & 0.995 & 0.754 & 0.998 & $1.69^{+0.17}_{-0.18}$ & P \\
\hline
1700.01* & 0.387 & 0.588 & 0.475 & $2.71^{+0.52}_{-0.52}$ & A \\ 
\hline
1899.01 & 0.535 & 0.582 & 0.616 & $4.38^{+0.94}_{-2.48}$ & A \\
1899.02* & 0.428 & 0.521 & 0.449 & $1.75^{+0.74}_{-0.83}$ & A \\
\hline
2289.01 & 0.166 & $>$0.999 & $>$0.999 & $2.25^{+0.51}_{-0.52}$ & P \\
2289.02* & $>$0.999 & $>$0.999 & $>$0.999 & $1.56^{+0.37}_{-0.38}$ & P \\
\hline
2580.01* & 0.457 & 0.585 & 0.542 & $2.17^{+0.71}_{-0.86}$ & A \\
\hline
2851.01 & 0.503 & 0.539 & 0.542 & $3.12^{+0.47}_{-0.49}$ & A \\
2851.02* & 0.497 & 0.510 & 0.506 & $2.05^{+0.34}_{-0.34}$ & A \\
\hline
3120.01* & 0.393 & 0.528 & 0.420 & $2.20^{+0.48}_{-0.51}$ & A \\
\hline
3401.01* & 0.999 & 0.989 & $>$0.999 & $1.60^{+0.32}_{-0.32}$ & P \\
3401.02 & 0.374 & 0.999 & 0.999 & $1.79^{+0.65}_{-0.65}$ & P \\
\hline
3456.01* & 0.459 & 0.503 & 0.462 & $1.61^{+0.28}_{-0.29}$ & A \\
3456.02 & 0.385 & 0.501 & 0.386 & $1.75^{+0.69}_{-0.69}$ & A \\
\hline
5845.01* & 0.502 & 0.497 & 0.499 & $1.68^{+0.49}_{-0.51}$ & A
\enddata
\tablecomments{The results for the sample of planets analyzed in this work. Column 1 contains the KOI numbers for the planets. Column 2 contains the Bayesian likelihood that the planet is orbiting the primary star using the density distribution overlaps. Column 3 contains the Bayesian prior probability that the planet is orbiting the primary star using our model of the single planet radius distribution. Column 4 contains the Bayesian posterior probability that the planet is orbiting the primary star. Column 5 contains the updated planet radius for the most probable host star case. Column 6 flags whether a planet is confidently circumprimary (P) or the host star is ambiguous (A). Planets with * are part of the radius gap subsample.}
\end{deluxetable*}

\section{Discussion}
\begin{figure*}[ht]
    \centering
    \includegraphics[angle=-90, width=0.65\textwidth]{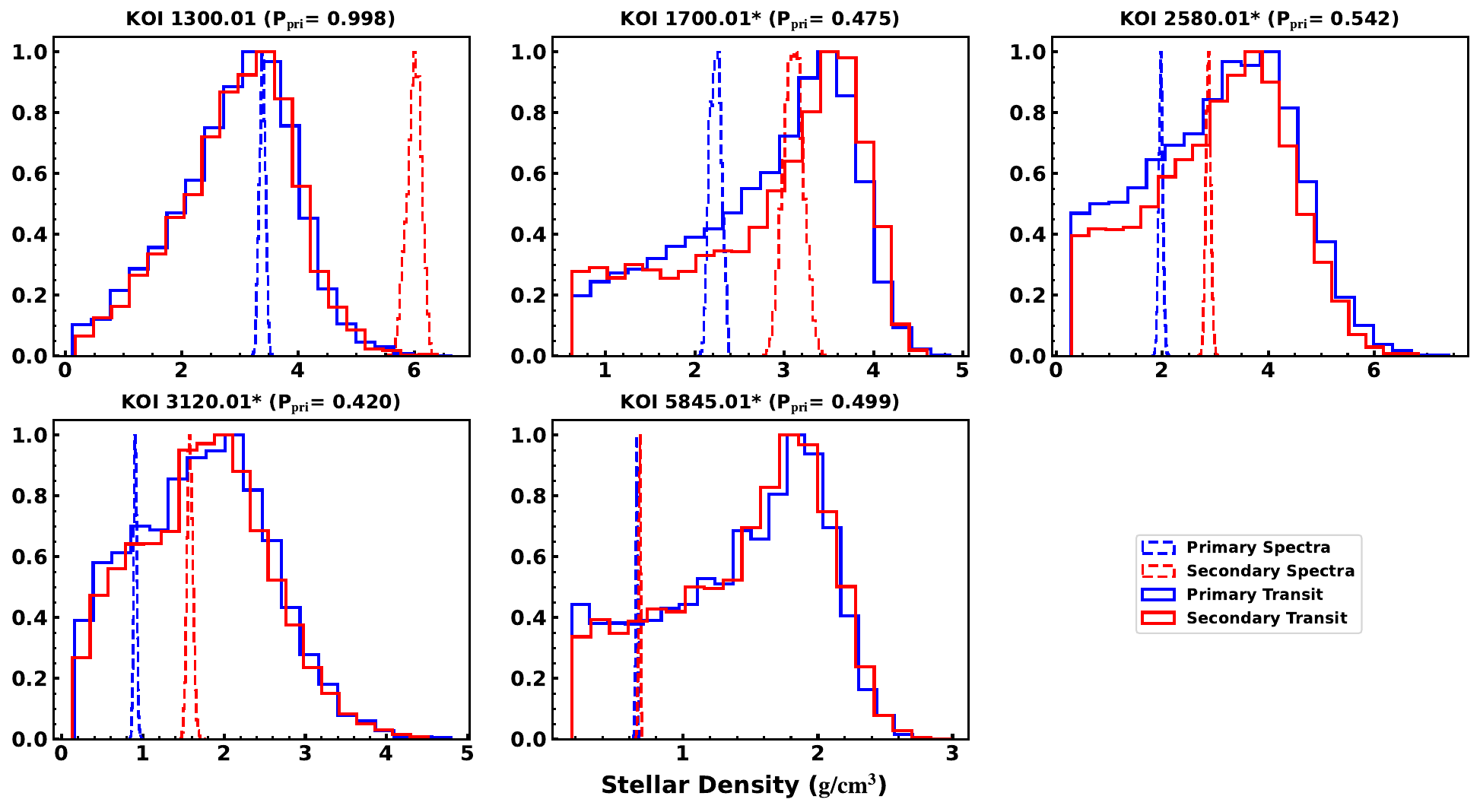}
    \caption{The density profiles for the 1-planet systems analyzed in this work. The solid blue histograms show the posterior density distribution from transit fitting for the primary star case. The solid red histograms show the same for the secondary star case. The dashed blue and red histograms show the \citet{Sullivan_2023} spectroscopic density distributions for the primary stars and secondary stars, respectively. Planets with a * in the title are in the radius gap subsample.}
    \label{fig:1planet}
\end{figure*}

\begin{figure*}[ht]
    \centering
    \includegraphics[angle=-90, width=0.8\textwidth]{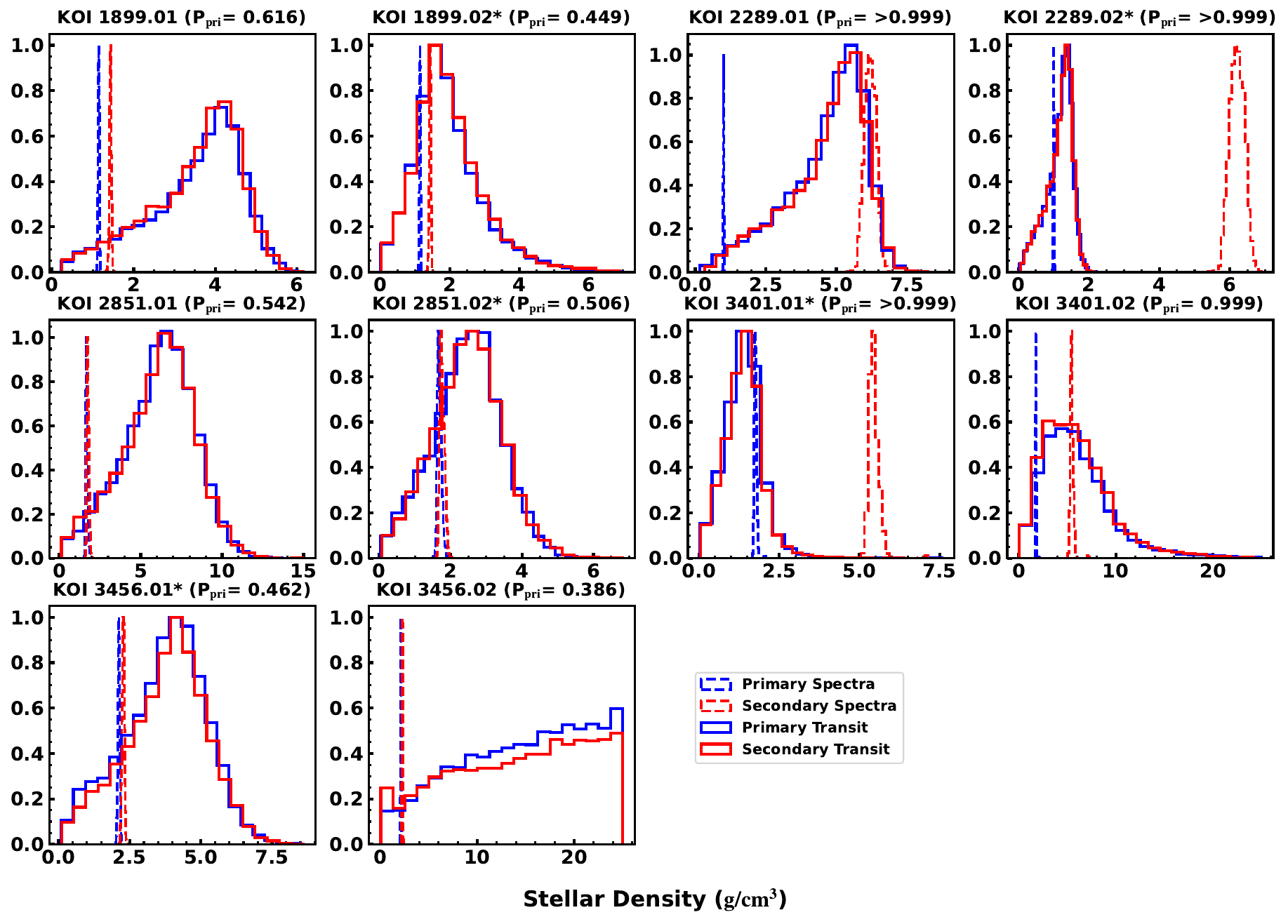}
    \caption{The density profiles for the 2-planet systems analyzed in this work. The solid blue histograms show the posterior density distribution from transit fitting for the primary star case. The solid red histograms show the same for the secondary star case. The dashed blue and red histograms show the \citet{Sullivan_2023} spectroscopic density distributions for the primary stars and secondary stars, respectively. Planets with a * in the title are in the radius gap subsample.}
    \label{fig:2planet}
\end{figure*}

The density posteriors for each fitted system are shown in Figures \ref{fig:1planet} and \ref{fig:2planet}. Based on the sums of the posterior probabilites, we found that $\sim66\%$ of planets are circumprimary and $\sim 34\%$ are circumsecondary. The density posterior distributions consistently have long tails toward lower densities and sharper cutoffs at high densities. This is largely due to covariance with the impact parameter $b$. The inferred stellar density, $\rho_\star$, is weakly sensitive to $b$ when $b$ is low, but as $b$ approaches 1, $\rho_\star$ decreases rapidly. Since the impact parameter is usually not well constrained in our sampling, this leads to a low density tail in most cases.

This systematic behavior is important because if the distribution shows more agreement with the primary star density, there will generally be little to no overlap with the secondary star density. Inversely, for distributions that shows more agreement with the secondary star density, the low density tail will usually overlap with the primary star density. This means that our likelihoods will either strongly favor a primary star host or weakly favor a secondary star host. Additionally, the prior that we are using favors planets with smaller radii. The planets would all have larger radii if they were hosted by the secondary star. This means that primary star hosts cannot be ruled out in general, but secondary star hosts can be ruled out. This is especially relevant in binaries where both stars have roughly similar densities, and it contributes to the large number of ambiguous results.

It is also important to note that because of the flux contamination in unresolved binaries, small planets are harder to detect around low-mass secondary stars. This means that the observed small planet population in unresolved binaries will be biased towards being circumprimary or in equal-mass binaries.

\begin{figure}
    \centering
    \gridline{\fig{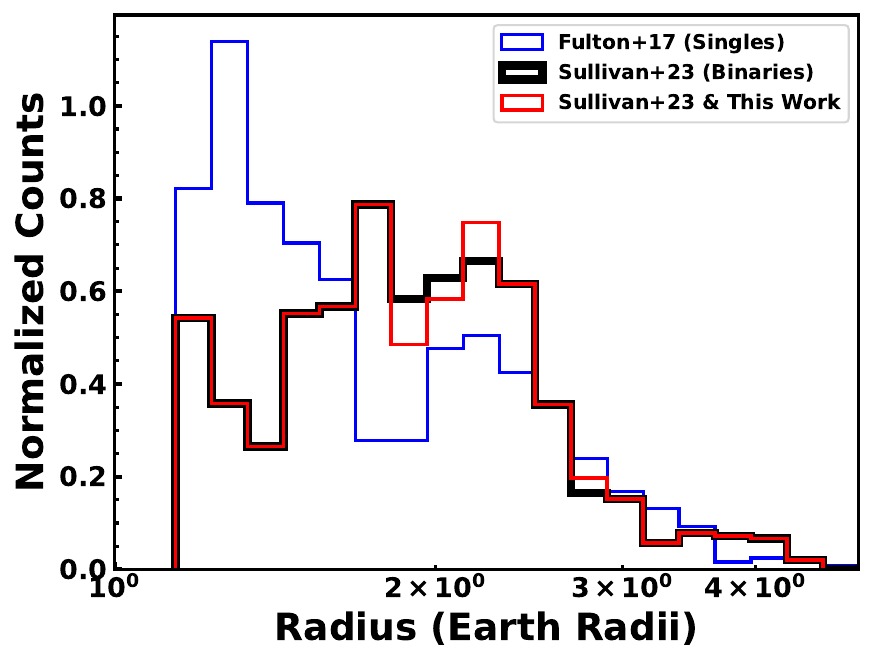}{0.45\textwidth}{}}
    \caption{The uncorrected radius distributions for planets in single and binary star systems. The thin blue distribution is the radius distribution for planets in single star star systems from \citet{Fulton_2017}. The black distribution is the circumprimary radius distribution for planets in binaries from \citet{Sullivan_2023}. The red distribution is the \citet{Sullivan_2023} circumprimary distribution modified to account for the more likely circumsecondary planets identified in this work. The change in the radius distribution for planets in binaries is minimal and tentatively confirms the absence of a radius gap reported in \citet{Sullivan_2023}.}
    \label{fig:new_rad_dist}
\end{figure}

Additionally, the posterior probabilities favored primary star hosts in most cases, suggesting that many of these planets likely do reside in the canonical radius gap, shown in Figure\textbf{ \ref{fig:new_rad_dist}}. This differs from results for single stars that have found a relatively empty radius gap \citep[e.g.][]{VanEylen_2018, Ho_2023}. \cite{Sullivan_2024} found that the radius gap for planets in binaries may still be present, but that the location of the gap shifts as a function of binary separation. The sample in this work is too small to investigate the role of binary separation.

\section{Conclusions}
In summary, our results provide preliminary confirmation that the canonical radius gap is well-populated among planets hosted in binary star systems, at least when binary separation is not accounted for. Nevertheless, our results also show the limitations of the asterodensity method for determining host stars, particularly in the case of near-equal mass binaries.

We only performed our analysis on 1 and 2-planet systems that satisfied our radius gap criteria. In the future, we will perform our analysis on the full sample of planets in binaries from \cite{Sullivan_2024} in order to construct a robust radius distribution for planets in binaries. With this we could determine if there is a radius gap for planets for binaries, including testing whether there is a separation dependence on the location of the gap. Beyond the radius gap question, if we performed our analysis on a large sample of planets, we could assemble statistics on the number of planets hosted by the primary star versus the secondary star. With a large sample of multi-planet systems, we could explore if planets typically all orbit the same star or some combination of both.

However, this work is fundamentally limited by the width of our density posterior distributions. We could obtain more narrow posterior distributions if our data had both a higher time cadence and higher signal-to-noise ratio. As discussed in Section \ref{30v60}, Kepler had a higher cadence (1 minute) mode, but the the lower signal-to-noise ratio balanced this out for our test case. Higher cadence alone is not sufficient. Data of this quality could potentially be obtained in the future by a telescope like the European Space Agency's PLAnetary Transits and Oscillations of stars (PLATO) telescope \citep{PLATO}.

Another way to achieve more conclusive results is to combine our asterodensity analysis with other techniques for identifying the host star. For example, we could look for shifts in the centroid of the KOIs in the Kepler imaging when the planet is in transit. If we had multi-band lightcurves of the transits, the color dependence of the transit depth would be consistent with the color of the host star, allowing the host star to be determined. Data of this kind could be acquired by future missions like ARIEL \citep{ARIEL}. In multi-planet systems, we could look for transit timing variations to confirm whether or not planets are hosted by the same star. Using these additional techniques could be particularly useful where both stars have similar densities, since the asterodensity technique is not effective in these cases.

Ultimately, using asterodensity profiling can be an effective method for determining the host stars of transiting exoplanets in unresolved binary star systems. Identifying the host star establishes more accurate planetary parameters. The asterodensity method is most effective when the binary contrast and signal-to-noise ratio of the transit are high. The degeneracy between the impact parameter of the transit and the host star density can pose a challenge to obtaining precise results. It is worth pursuing future work to obtain a statistical sample of planets analyzed with this method.

\section{Acknowledgments}
This paper includes data collected by the Kepler mission and obtained from the MAST data archive at the Space Telescope Science Institute (STScI) \citep{MAST_DR25}. Funding for the Kepler mission is provided by the NASA Science Mission Directorate. STScI is operated by the Association of Universities for Research in Astronomy, Inc., under NASA contract NAS 5–26555. This research has made use of the NASA Exoplanet Archive, which is operated by the California Institute of Technology, under contract with the National Aeronautics and Space Administration under the Exoplanet Exploration Program. The authors acknowledge the Texas Advanced Computing Center (TACC) at The University of Texas at Austin for providing computational resources that have contributed to the research results reported within this paper. NBW received support for the initial stages of this program from NSF REU grant AST-2244278 (PI: Jogee). NBW was funded in part by the UT-Austin Center for Planetary Systems Habitability.

\bibliography{references}{}
\bibliographystyle{aasjournal}

\end{document}